\documentclass[aps,prd,twocolumn,nofootinbib,superscriptaddress,10pt]{revtex4-1}
\usepackage{amsmath,amsthm,latexsym,amssymb,amsfonts}
\usepackage{graphicx,color}
\usepackage{hyperref,natbib}
\usepackage[utf8]{inputenc}
\usepackage[T1]{fontenc}
\usepackage{textcomp}

\begin{document}

\title{Hamiltonian formulation of teleparallel gravity}
\author{Rafael Ferraro}
\email{ferraro@iafe.uba.ar}
\thanks{member of Carrera del Investigador Cient\'{\i}fico (CONICET,
Argentina). }\affiliation{Instituto de Astronom\'\i a y F\'\i sica
del Espacio (IAFE, CONICET-UBA), Casilla de Correo 67, Sucursal 28,
1428 Buenos Aires, Argentina.} \affiliation{Departamento de F\'\i
sica, Facultad de Ciencias Exactas y Naturales, Universidad de
Buenos Aires, Ciudad Universitaria, Pabell\'on I, 1428 Buenos Aires,
Argentina.}

\author{Mar\'\i a Jos\'e Guzm\'an}
\email{mjguzman@iafe.uba.ar}\affiliation{Instituto de Astronom\'\i
a y F\'\i sica del Espacio (IAFE, CONICET-UBA), Casilla de Correo
67, Sucursal 28, 1428 Buenos Aires, Argentina.}

\begin{abstract}
{\ The Hamiltonian formulation of the teleparallel equivalent of
general relativity (TEGR) is developed from an ordinary second-order
Lagrangian, which is written as a quadratic form of the coefficients
of anholonomy of the orthonormal frames (vielbeins). We analyze the
structure of eigenvalues of the multi-index matrix entering the
(linear) relation between canonical velocities and momenta to obtain
the set of primary constraints. The canonical Hamiltonian is then
built with the Moore-Penrose pseudo-inverse of that matrix. The set
of constraints, including the subsequent secondary constraints,
completes a first class algebra. This means that all of them
generate gauge transformations. The gauge freedoms are basically the
diffeomorphisms, and the (local) Lorentz transformations of the
vielbein. In particular, the ADM algebra of general relativity is
recovered as a sub-algebra.}
\end{abstract}

\maketitle

\section{Introduction}

The determination of the independent dynamical degrees of freedom is of the
utmost importance in any field theory, since it allows to exhibit the
internal consistency of the theory, and tackle the issue of the
well-posedness of the Cauchy problem. It also puts the theory into a
different perspective, because it helps to find the minimal number of
variables specifying the state of the system, so being vital for the
quantization of the theory. According to the procedure due to Dirac \cite
{Dirac1964}, the number of genuine degrees of freedom can be determined from
the algebra of the constraints among the canonical variables of the theory.
The constraints firstly appear when the canonical momenta are computed.
These \textit{primary} constraints have to be consistent with the
Hamiltonian evolution of the system, which leads to \textit{secondary}
constraints, and so on. Finally, the set of all the constraints is
reclassified as \textit{first class} and \textit{second class }constraints,
depending whether their Poisson brackets are or not null on the constraint
surface in the phase space. First class constraints generate gauge
transformations; so, each of them is related to a spurious degree of
freedom. On the other hand, second class constraints can be reorganized as
pairs of spurious conjugated variables. Thus, the number of genuine degrees
of freedom can be computed as
\begin{eqnarray}
\#\,\text{d.o.f.}\ &=&\ \#\,\text{pairs of canonical variables}  \notag \\
&&-\#\,\text{first class constraints}  \notag \\
&&-\frac{1}{2}\ \#\,\text{second class constraints}\ .  \label{dof}
\end{eqnarray}
A nice example is the Maxwell potential, described by four dynamical
variables $A_{\mu }$ that are governed by the Lagrangian $L[A_{\mu }]\propto
F_{\lambda \rho }\ F^{\lambda \rho }$ (the field tensor $F_{\lambda \rho }$
is $F_{\lambda \rho }=\partial _{\lambda }A_{\rho }-\partial _{\rho
}A_{\lambda }$). Since $F_{\lambda \rho }$ is anti-symmetric, then $\partial
_{0}A_{0}$ is not present in the Lagrangian. Thus the canonical momentum $
\pi ^{0}=\partial L/\partial (\partial _{0}A_{0})$ identically vanishes; it
is a primary constraint. The consistency of the constraint $\pi ^{0}=0$ with
the evolution requires the vanishing of the Poisson bracket between $\pi
^{0} $ and the Hamiltonian; this leads to the secondary constraint $\nabla
_{i}\pi ^{i}\propto \nabla _{i}F^{0i}=0$ (Gauss's law). Both constraints are
first class, since the Poisson brackets between canonical momenta are
identically null. Therefore, according to Eq.~(\ref{dof}), one realizes that
the electromagnetic field has not four degrees of freedom $A_{\mu }$ at each
event, but only two (electromagnetic waves are transversal). At the level of
the initial data, the existence of constraints imply a restriction on the
spectrum of allowed initial configurations. Besides, the absence of kinetic
term for $A_{0}$ in the Lagrangian implies that the evolution of this
dynamical variable, conjugate to the first class constraint $\pi ^{0}$,
remains completely undetermined. The same happens to the evolution of the
longitudinal component of the potential $A_{||}$, which also remains
undetermined as a consequence of the existence of the first class constraint
$\nabla _{i}\pi ^{i}$. Thus, $A_{0}$ and $A_{||}$ are gauge freedoms.\ The
former conclusions can also be derived from a slightly modified Lagrangian.
The integration by parts of one of the terms containing $\partial _{i}A_{0}$
leads to a surface term, which can be eliminated, plus the term $
A_{0}\,\nabla _{i}F^{0i}$. In such way, the spurious degree of freedom $
A_{0} $ becomes a Lagrange multiplier whose variation leads to the Gauss's
law constraint (any other presence of $A_{0}$ is captured in the canonical
momenta $\pi ^{i}$) \cite{Sundermeyer1982}.

The canonical formulation of general relativity (GR) relies on the widely
spread formalism by Arnowitt, Deser and Misner (ADM) \cite{Arnowitt:1962hi},
in which the spacetime is foliated into a family of spacelike hypersurfaces
that induces a proper decomposition of the metric tensor $g_{\mu \nu }$. The
Einstein-Hilbert Lagrangian can be integrated by parts to realize that the
temporal sector of the metric (the \textit{lapse} $N$ and the \textit{shift
vector }$N_{i}$) is thrown into the role of Lagrange multipliers associated
to four first class constraints (the super-Hamiltonian and super-momenta
constraints). So written, the Lagrangian gives dynamics only to the six
components of the 3-dimensional metric $g_{ij}$ on the spacelike
hypersurfaces of the foliation; but the canonical variables $(g_{ij},\,\pi
^{ij})$ are still constrained by the four first class constraints. Thus the
gravitational field contains only two genuine degrees of freedom. In fact,
apart from the undetermined evolutions of the four Lagrange multipliers $
(N,\,N_{i})$, there are also four gauge freedoms among the six components of
$g_{ij}$ (gravitational waves are transversal and trace-less). As a feature
that distinguishes GR from electromagnetism, the GR Hamiltonian vanishes
because of the constraints. This feature is typical of systems having a time
hidden among their canonical variables \cite{Kuchar1992}.

Early in the 1918, Weyl's unsuccessful attempt of unifying
gravitation and electromagnetism introduced for the first time the
notion of gauge theories \cite{Weyl1918}. Einstein himself tried ten
years later the same unification idea, but taking advantage of the
sixteen components of the tetrad field in order to include the
electromagnetic field \cite{Einstein_1925}. Later he realized that
the arbitrariness in the choice of the tetrad comes from the set of
local Lorentz transformations that leave the metric unchanged,
therefore the extra degrees of freedom could not give account for
electromagnetism. However, he introduced the concepts of
teleparallelism that remain important until today, presenting for
the first time the teleparallel equivalent of general relativity
(TEGR), an equivalent formulation of general relativity. In fact,
although both theories have different Lagrangian formulations, they
are equivalent at the level of the equations of motion. Nonetheless,
they are based on completely different Lagrangian constructions.
This is so because TEGR describes gravity as the effect of torsion
in the curvatureless Weitzenb\"{o}ck geometry; the dynamical variables
are not the components of the metric $g_{\mu \nu }$ but those of the
field of orthonormal frames --tetrads or vierbeins-- $e_{\mu }^{a}$
($a$ and $\mu $ are $SO(3,1)$ and coordinate indices, respectively)
\cite{Einstein_1925,Cartan_1922}. As a consequence, the Hamiltonian
formalisms of GR and TEGR are different too. Among the articles
treating the Hamiltonian formulation of TEGR we specially mention
Ref.~\cite{Maluf_2001}, which introduces a set of auxiliary
variables in a first order approach that
lowers the order of the Euler-Lagrange equations (cf. 
\cite{Maluf_1994,Maluf_1999,da_Rocha_Neto_2010,Maluf_2013}), and
Ref.~\cite{Blagojevic_2000} that deals with an enlarged set of
variables and constraints to enforce the vanishing of the curvature.
The canonical formulation of TEGR has been also stated in the
geometric language of differential forms
\cite{Okolow_2013,Okolow_2014}.

In this work we will put forward the Hamiltonian formalism for TEGR in a way
as close as possible to the second order formalism of electrodynamics that
was sketched above. This work is organized as follows: in Section \ref{standard_tegr} 
we introduce the standard TEGR dynamics, which is governed
by a Lagrangian quadratic in the torsion. In Section \ref{new_tegr} we show
that the TEGR Lagrangian can be reformulated as the quadratic inner product
of the anholonomy coefficients with respect to a supermetric that is defined
in the tangent space. In Section \ref{H_and_Pi} we obtain the set of primary
and secondary constraints that are equivalent to those of electrodynamics
and GR geometrodynamics. In Section \ref{gauge_t} we study the gauge
transformations generated by these constraints (they will prove to be first
class). Compared with geometrodynamics, TEGR has an additional gauge
symmetry associated to local Lorentz transformations of frames, which is the
source of the constraints analyzed in Section \ref{lorentz_g}. In Section 
\ref{canonical_H} the (constrained) linear relations between canonical
momenta and velocities is inverted to build the canonical TEGR Hamiltonian $
\mathcal{H}$; the procedure implies a careful analysis of the eigenvector
structure involved in these linear relations, in order to build the
respective pseudo-inverse matrix. The entire set of $n(n+3)/2$ constraints ($
n$ is the spacetime dimension) is consistent with the evolution governed by $
\mathcal{H}$; besides, they are first class as proven by the algebra of
constraints computed in Section \ref{full_brackets}. In Section \ref{conclusions} 
we summarize the main steps and the achievements of
the paper. The Appendix \ref{app} shows some useful computations
that are needed throughout the work.

\section{TEGR and standard Lagrangian formulation}

\label{standard_tegr}

TEGR is a theory of gravity where the field of orthonormal frames plays the
role of dynamical variable. Let $M$ be a manifold, $\{\mathbf{e}_{a}\}$ a
basis in the tangent space $T_{p}(M)$, and $\{\mathbf{E}^{a}\}$ its dual
basis in the cotangent space $T_{p}^{\ast }(M)$ (i.e., if the 1-forms $
\mathbf{E}^{a}$ are applied to the vectors $\mathbf{e}_{b}$ one obtains $
\mathbf{E}^{a}(\mathbf{e}_{b})=\delta _{b}^{a}$). They can be expanded in a
coordinate basis as $\mathbf{e}_{a}=e_{a}^{\mu }~\partial _{\mu }$ and $
\mathbf{E}^{a}=E_{\mu }^{a}\ dx^{\mu }$; so duality means that
\begin{equation}
E_{\mu }^{a}\ e_{b}^{\ \mu }~=~\delta _{b}^{a}~,\ \ \ \ \ e_{a}^{\mu
}~E_{\nu }^{a}~=~\delta _{\nu }^{\mu }~.  \label{duality}
\end{equation}
Here and from now on, we will use Greek letters $\mu ,\nu ,...=0,...,n-1$
for spacetime coordinate indices, and Latin letters $a,b,...,g,h=0,...,n-1$
for Lorentzian tangent space indices. A \textit{vielbein} (vierbein o tetrad
in $n=4$ dimensions) is a basis encoding the metric structure of the
spacetime:
\begin{equation}
\mathbf{g~}=~\eta _{ab}\ \mathbf{E}^{a}\otimes \mathbf{E}^{b}~,
\label{metric1}
\end{equation}
therefore,
\begin{equation}
\mathbf{E}^{a}\cdot \mathbf{E}^{b}~=~\mathbf{g}(\mathbf{\mathbf{E}}^{a},
\mathbf{\mathbf{E}}^{b})\mathbf{~}=~\eta _{ab}~,
\end{equation}
which means that the vielbein is an orthonormal basis. In component
notation, the former expressions look
\begin{equation}
g_{\mu \nu }~=~\eta _{ab}~E_{\mu }^{a}~E_{\nu }^{b}~,\hspace{0.25in}\eta
_{ab}~=~g_{\mu \nu }~e_{a}^{\mu }~e_{b}^{\nu }~,  \label{metric2}
\end{equation}
which implies that the relation between the metric volume and the
determinant of the matrix $E_{\mu }^{a}$ is
\begin{equation}
\sqrt{|g|}~=~\text{det}[E_{\mu }^{a}]~\doteq ~E~.  \label{volume}
\end{equation}

Since the vielbein encodes the metric structure of the spacetime, one can
formulate a dynamical theory of the spacetime geometry by defining a
Lagrangian for the vielbein field. In particular, there is a Lagrangian
which leads to dynamical equations for the vielbeins that are equivalent to
Einstein equations for the metric \cite{Hayashi1979}. The so called
teleparallel equivalent of general relativity is governed by the
Lagrangian density
\begin{equation}
L~=~E\ T~,  \label{TEGR}
\end{equation}
where $T$ is the torsion scalar
\begin{equation}
T~\doteq ~T_{\ ~\mu \nu }^{\rho }\ S_{\rho }^{\ ~\mu \nu },  \label{Tscalar}
\end{equation}
which is made up of
\begin{equation}
T_{\hspace{0.05in}\,\nu \rho }^{\mu }~\doteq ~e_{a}^{\ \mu }~(\partial _{\nu
}E_{\rho }^{a}-\partial _{\rho }E_{\nu }^{a})~,  \label{TorsionTensor}
\end{equation}
and
\begin{equation}
S_{\rho }^{\ \mu \nu }~\doteq ~\dfrac{1}{2}\left( K_{\hspace{0.05in}\hspace{
0.05in}\rho }^{\mu \nu }+T_{\lambda }^{\ \lambda \mu }~\delta _{\rho }^{\nu
}-T_{\lambda }^{\ \lambda \nu }~\delta _{\rho }^{\mu }\right) ~,
\end{equation}
where
\begin{equation}
K_{\hspace{0.05in}\hspace{0.05in}\rho }^{\mu \nu }~\doteq ~\frac{1}{2}
(T_{\rho }^{\hspace{0.05in}\mu \nu }-T_{\hspace{0.05in}\hspace{0.05in}\rho
}^{\mu \nu }+T_{\hspace{0.05in}\hspace{0.05in}\rho }^{\nu \mu })~.
\label{contorsion}
\end{equation}
In Lagrangian (\ref{TEGR}), the strength field $T_{\ ~\nu \rho
}^{\mu }$ is the torsion associated with the Weitzenb\"{o}ck connection
$\Gamma _{\nu \rho }^{\mu }\doteq e_{a}^{\ \mu }\,\partial _{\nu
}E_{\rho }^{a}$, and $K_{\hspace{0.1in}\rho
}^{\mu \nu }$ is the contorsion \cite{Ortin}. In geometric language, torsion is the 2-form $\mathbf{T}^{a}\doteq d
\mathbf{E}^{a}+\mathbf{\omega }_{\hspace{0.05in}b}^{a}\wedge \mathbf{E}^{b}$
, where the 1-form $\mathbf{\omega }_{\hspace{0.05in}b}^{a}$ is the spin
connection. Weitzenb\"{o}ck connection is the choice $\mathbf{\omega }\,_{
\hspace{0.05in}b}^{a}=0$, because it leads to $(\mathbf{T}^{a})_{\nu \rho
}=(d\mathbf{E}^{a})_{\nu \rho }=\partial _{\nu }E_{\rho }^{a}-\partial
_{\rho }E_{\nu }^{a}=E_{\mu }^{a}~T_{\hspace{0.05in}\nu \rho }^{\mu }$.
Weitzenb\"{o}ck connection is metric compatible, since $\nabla _{\nu }E_{\mu
}^{a}=\partial _{\nu }E_{\mu }^{a}-\Gamma \,_{\nu \mu }^{\lambda }E_{\lambda
}^{a}=0$. Besides, from Eq.~(\ref{duality}) we also get that $\nabla _{\nu
}e_{a}^{\mu }=0$. This means that the vielbein is automatically
parallel-transported along any curve. Furthermore, the parallel-transport of
any vector does not depend on the path (it is \textit{absolute}), since
Weitzenb\"{o}ck connection has the remarkable feature that the curvature $
\mathbf{R}_{\hspace{0.05in}b}^{a}\doteq d\mathbf{\omega }_{\hspace{0.05in}
b}^{a}+\mathbf{\omega }_{\hspace{0.05in}c}^{a}\wedge \mathbf{\omega }_{
\hspace{0.05in}b}^{c}$ is identically zero. The (Weitzenb\"{o}ck) covariant
derivative of a vector is $\nabla _{\nu }\mathbf{U}=\nabla _{\nu }(U^{a}
\mathbf{e}_{a})=\mathbf{e}_{a}~\partial _{\nu }U^{a}$; thus, vector
$\mathbf{U}$ will be parallel transported if and only if its
components $U^{a}$ are constant.

Although TEGR Lagrangian can be understood in terms of the
Weitzenb\"{o}ck connection and its respective torsion, it should be
emphasized that the TEGR Lagrangian neither fixes the connection nor
the vielbein; it only determines the metric, as it is well known.
Furthermore, whenever matter couples minimally to the metric, as
usual, the free particles will follow geodesics of the (torsionless)
Levi-Civita connection $\overline{\Gamma }_{\nu \rho }^{\mu }$.
\footnote{However, Levi-Civita and Weitzenb\"{o}ck connections are
related through the contorsion: $\overline{\Gamma }_{\nu \rho }^{\mu
}={\Gamma }_{\nu \rho }^{\mu }-K_{\hspace{0.05in}\nu\rho
}^{\mu}\,$.} Setting aside this point, we use to say that TEGR is a
theory where the gravitational effects are fully encoded in the
torsion. On the contrary, GR associates gravity to curvature; it
assumes that the spacetime is endowed with the torsionless
Levi-Civita connection, whose curvature enters the Einstein-Hilbert
Lagrangian $L=E~\overline{R}$. The reason why TEGR is indeed
equivalent to GR is traced to the fact that their respective
Lagrangian densities differ in a surface term:
\begin{equation}
-E\ \overline{R}=E\ T-2\ \partial _{\rho }(E\ T_{\mu }^{\ \mu \rho }),
\label{RLC}
\end{equation}
Even so the vielbein field contains $n^{2}$ components, while the metric
tensor has only $n(n+1)/2$. However, TEGR dynamical equations are invariant
under local Lorentz transformations of the vielbein, which involve $\binom{n
}{2}$ generators. Such a gauge invariance means that $\binom{n}{2}=n(n-1)/2$
degrees of freedom cancels out, which allows that the theories turn out to
be equivalent at the level of the equations of motion.

\section{TEGR Lagrangian in terms of the vielbein field}

\label{new_tegr}

With the aim of preparing the TEGR Lagrangian for the study of its canonical
structure, we will rewrite it completely in terms of $e_{a}^{\mu }$, $E_{\nu
}^{a}$ and the derivatives $\partial _{\mu }E_{\nu }^{a}$. This imply the
removing of any presence of the metric field, since such contributions hide
a dependence on the vielbein. We transform the scalar torsion into
\begin{equation}
T~~=~\frac{1}{4}~T_{\rho }^{\ \,\mu \nu }~T_{\,\,\mu \nu }^{\rho }~-~\frac{1
}{2}~T_{\,\,\mu \nu }^{\rho }~T_{\,\,\,\,\,\,\rho }^{\mu \nu
}~-~T_{\,\,\,\,\mu \rho }^{\rho }~T_{\,\,\,\,\,\,\,\,\nu }^{\nu \mu }~.
\end{equation}
We note that all terms in $T$ are quadratic in the antisymmetrized
derivatives of the vielbein; writing term by term one gets
\begin{equation}
\frac{1}{4}~T_{\rho }^{\ \,\mu \nu }~T_{\,\,\mu \nu }^{\rho }~=~\dfrac{1}{4}
~g_{\rho \alpha }~g^{\beta \mu }~g^{\gamma \nu }~T_{\,\,\beta \gamma
}^{\alpha }~T_{\,\,\mu \nu }^{\rho }~;
\end{equation}
then one replaces the expressions for the torsion tensor \ref{TorsionTensor} and the metric in
terms of the vielbein field and its inverse \ref{metric2}:
\begin{equation}
\frac{1}{4}\,T_{\rho }^{\ \,\,\mu \nu }\,T_{\,\,\mu \nu }^{\rho
}=\eta _{ab}\,\eta ^{c[d}\,\eta ^{f]e}\,E\ \partial _{\mu
}E_{\,\,\nu }^{a}\ \partial _{\rho }E_{\,\,\lambda }^{b}\ e_{c}^{\mu
}\ e_{e}^{\nu }\ e_{d}^{\rho }\ e_{f}^{\lambda }\,.\ \
\end{equation}
After this procedure has been performed in all the terms, the TEGR
Lagrangian becomes
\begin{equation}
L=E\ T=\frac{1}{2}~E\ \partial _{\mu }E_{\,\,\nu }^{a}\ \partial _{\rho
}E_{\,\,\lambda }^{b}\ e_{c}^{\mu }~e_{e}^{\nu }~e_{d}^{\rho
}~e_{f}^{\lambda }\,M_{ab}^{\ \,\,\,\,cedf}\,,  \label{LagrM}
\end{equation}
where we call \emph{supermetric} $M_{ab}^{\ \,\,\,cedf}$ the emerging
Lorentz invariant tensor given by
\begin{equation}
M_{ab}^{\ \,\,\,cedf}\doteq 2\,\eta _{ab}\,\eta ^{c[d}\,\eta
^{f]e}-4\,\delta _{a}^{[d}\,\eta ^{f][c}\,\delta _{b}^{e]}+8\,\delta
_{a}^{[c}\,\eta ^{e][d}\,\delta _{b}^{f]}\,.  \label{supermetric}
\end{equation}
The supermetric is antisymmetric in the pairs of indices $c-e$ and $d-f$,
what implies that only the antisymmetric parts of $\partial _{\mu
}E_{\,\,\nu }^{a}$ and $\partial _{\lambda }E_{\,\,\rho }^{b}$ take part in
the Lagrangian (\ref{LagrM}).
Other properties of the supermetric are summarized in the Appendix \ref{smprop}. \newline

We remark that the index structure of the supermetric is natural when we
recognize in Eq.~(\ref{LagrM}) the anholonomy coefficients $f_{ab}^{c}$,
which are defined by the commutator $[\mathbf{e}_{a},\mathbf{e}
_{b}]=f_{ab}^{c}~\mathbf{e}_{c}$. In fact, by using the equations (\ref{duality}) the coefficients $f_{bc}^{a}$ can be rewritten as
\begin{equation}
f_{bc}^{a}~=~-e_{b}^{\mu }~e_{c}^{\nu }~(\partial _{\mu }E_{\nu
}^{a}-\partial _{\nu }E_{\mu }^{a})~=~-2~e_{b}^{\mu }~e_{c}^{\nu }~\partial
_{\lbrack \mu }E_{\nu ]}^{a}~,  \label{Anhol}
\end{equation}
which can be related to other geometrical magnitudes, as the Weitzenb\"{o}ck
torsion and the Lie derivative of the vielbein:
\begin{equation}
f_{bc}^{a}~=~\mathbf{T}^{a}(\mathbf{e}_{c},~\mathbf{e}_{b})~=~(\mathcal{L}_{\mathbf{e}_{c}}\mathbf{E}^{a})(\mathbf{e}_{b})~.
\end{equation}
In terms of these coefficients, the Lagrangian density looks in a very
elegant form:
\begin{equation}
L~=~\dfrac{1}{8}~E~f_{ce}^{a}~f_{df}^{b}~M_{ab}^{\ \,\,\,cedf}~.
\label{Lagrf}
\end{equation}
A similar expression for the Lagrangian can be found in Ref. \cite{Cho1976},
where the anholonomy coefficients are identified with a Yang-Mills-like
field strength; however, that Lagrangian still mixed tangent space and
coordinate indices. Instead, Lagrangian (\ref{Lagrf}) does not involve
coordinate indices; it shows that supermetric $M_{ab}^{\ \,\,\,cedf}$ is a
relevant geometric object in the (co-) tangent space structure of the
spacetime. We intend to analyze the Hamiltonian structure of TEGR by
starting from Lagrangian (\ref{LagrM}, \ref{Lagrf}), and following a
canonical second-order procedure.

\section{Super-Hamiltonian and super-momenta constraints}

\label{H_and_Pi}

We compute the canonical momenta by differentiating the Lagrangian (\ref{LagrM}) 
with respect to the time derivative of the canonical variable $E_{\mu }^{a}$:
\begin{eqnarray}
\Pi _{a}^{\mu } &=&\frac{\partial L}{\partial (\partial _{0}E_{\mu }^{a})}
=E\ \partial _{\rho }E_{\ \lambda }^{b}\ e_{c}^{0}~e_{e}^{\mu }~e_{d}^{\rho
}~e_{f}^{\lambda }\ M_{ab}^{\ \,\,\,cedf}  \notag \\
&=&-\frac{1}{2}~E~e_{c}^{0}~e_{e}^{\mu }~f_{df}^{b}~M_{ab}^{\ \,\,\,cedf}~.
\label{momfull}
\end{eqnarray}
So, the Poisson brackets in TEGR are defined as
\begin{eqnarray}
&&\{A(t,\mathbf{x}),\ B(t,\mathbf{y})\}\ \doteq  \notag \\
&&\ \ \int d\mathbf{z}\ \left( \dfrac{\delta A(t,\mathbf{x})}{\delta
E_{\lambda }^{a}(\mathbf{z})}\dfrac{\delta B(t,\mathbf{y})}{\delta \Pi
_{a}^{\lambda }(\mathbf{z})}-\dfrac{\delta A(t,\mathbf{x})}{\delta \Pi
_{a}^{\lambda }(\mathbf{z})}\dfrac{\delta B(t,\mathbf{y})}{\delta E_{\lambda
}^{a}(\mathbf{z})}\right) \ .
\end{eqnarray}
The brackets between fundamental canonical variables are
\begin{equation}
\{E_{\mu }^{a}(t,\mathbf{x}),~\Pi _{b}^{\nu }(t,\mathbf{y})\}\ =~\delta
_{b}^{a}~\delta _{\mu }^{\nu }~\delta (\mathbf{x}-\mathbf{y})~.
\end{equation}
Additional fundamental Poisson brackets, including $E$, $e_{a}^{\mu }$,
etc., are summarized in Appendix \ref{pbrackets}.

From Eq.~(\ref{momfull}) we immediately get $n$ trivial primary constraints
\begin{equation}
G_{a}^{(1)}~\doteq ~\Pi _{a}^{0}~\equiv ~0~,  \label{G1a}
\end{equation}
which are derived by noticing that $e_{c}^{0}~e_{e}^{0}$ is symmetric in $c-e$ 
but $M_{ab}^{\ \,\,\,cedf}$ is antisymmetric. Although we cannot prove
yet that they are first class (i.e. we do not know yet whether they generate
gauge transformations), the electromagnetic analogue tells us that they mean
the $E_{0}^{a}$'s are spurious gauge dependent variables, that would become
Lagrange multipliers if an integration by parts were performed in the
action. This is in line with the spurious character of the temporal sector
of the metric tensor we have commented in Section I.

The primary constraints must be satisfied at any time. In other words, if
the system is on the constraint surface at the initial time, it must remain
there along the evolution. If this consistency requirement were not
accomplished, then it could be enforced by resorting to new (secondary)
constraints \cite{Anderson_1951}. From a Hamiltonian perspective, the
consistency of the primary constraints is controlled by means of the primary
Hamiltonian \cite{Sundermeyer1982}
\begin{equation}
H_{p}~=~H+\int d\mathbf{x}~u^{a}(t,\mathbf{x})~\phi _{a}^{(1)}(t,\mathbf{x}
)~,  \label{HP}
\end{equation}
where $H$ is the canonical Hamiltonian, $u^{a}(t,\mathbf{x})$ are arbitrary
functions, and $\phi _{a}^{(1)}$ are \textit{all} the primary constraints.
The consistency will be fulfilled if the Poisson brackets $\{\phi
_{a}^{(1)},H_{p}\}$ are null on the constraint surface. This requirement
could be satisfied by properly choosing the functions $u^{a}(t,\mathbf{x})$;
if not, new (secondary) constraints will be needed to enforce it, and so on.
Actually, in TEGR we will find that all the Poisson brackets between
constraints are null on the constraint surface. This means that primary and
secondary constraints are all first class; they generate gauge
transformations. Thus the constraints will be consistent with the evolution
if their Poisson brackets with $H$ vanish on the constraint surface (i.e.,
if $H$ is gauge invariant, as it should be expected).\ \ \ \ \ \ \ \ \ \ \ \
\ \ \ \ \ \ \ \

\bigskip

In spite of the entire set of primary constraints was not obtained yet, the
evolution of constraints (\ref{G1a}) can be analyzed at the level of the
Euler-Lagrange evolution equations,
\begin{equation}
\partial _{\mu }\frac{\partial L}{\partial (\partial _{\mu }E_{\nu }^{a})}-
\frac{\partial L}{\partial E_{\nu }^{a}}~=~0~.  \label{E-L}
\end{equation}
By splitting the first term, one gets
\begin{equation}
\partial _{0}\Pi _{a}^{\nu }+\partial _{i}\frac{\partial L}{\partial
(\partial _{i}E_{\nu }^{a})}-\frac{\partial L}{\partial E_{\nu }^{a}}~=~0~.
\label{MOV}
\end{equation}
Therefore, if the constraints (\ref{G1a}) must be fulfilled at any time, we
obtain $n$ equations -- those having $\nu =0$ -- which do not contain
second-order temporal derivatives:
\begin{equation}
\partial _{i}\frac{\partial L}{\partial (\partial _{i}E_{0}^{a})}-\frac{
\partial L}{\partial E_{0}^{a}}~=~0~.
\end{equation}
Like the Gauss's law in electromagnetism these equations do not contain
dynamics, but they constrain the dynamics. Since the derivatives of the
vielbein enter the Lagrangian only in antisymmetric combinations, then it is
\begin{equation}
\partial _{i}\frac{\partial L}{\partial (\partial _{i}E_{0}^{a})}
~=~-\partial _{i}\frac{\partial L}{\partial (\partial _{0}E_{i}^{a})}
~=~-\partial _{i}\Pi _{a}^{i}~.
\end{equation}
Thus, we have found $n$ secondary constraints:
\begin{equation}
\partial _{i}\Pi _{a}^{i}+\frac{\partial L}{\partial E_{0}^{a}}~=~0~.
\label{secondary}
\end{equation}
We will prove that these constraints are consistent with the evolution; so
they do not generate new constraints. For this, we will apply the derivative
$\partial _{0}$ to the constraints (\ref{secondary}), and use the Eq.~(\ref{MOV}) 
to replace $\partial _{0}\Pi _{a}^{i}$:
\begin{eqnarray}
&&\partial _{0}\left( \partial _{i}\Pi _{a}^{i}+\frac{\partial L}{\partial
E_{0}^{a}}\right) =-\partial _{i}\partial _{j}\frac{\partial L}{\partial
(\partial _{j}E_{i}^{a})}+\partial _{\mu }\frac{\partial L}{\partial E_{\mu
}^{a}}~  \notag \\
&=&-\partial _{i}\partial _{j}\frac{\partial L}{\partial (\partial
_{j}E_{i}^{a})}+\partial _{\nu }\partial _{\mu }\frac{\partial L}{\partial
(\partial _{\mu }E_{\nu }^{a})}\equiv 0
\end{eqnarray}
(we also use Eq.~(\ref{E-L}) in the last step). The null result comes from
the fact that $\partial _{\mu }E_{\nu }^{a}$ enters the Lagrangian in
antisymmetric combinations but the operators $\partial _{i}\partial _{j}$
and $\partial _{\nu }\partial _{\mu }$ are symmetric.

So far, we have got a set of constraints which is consistent with the
evolution. To write them in a fully canonical way, we have to compute the
derivative $\partial L/\partial E_{0}^{a}$ and express it as a function of
the momenta, the vielbein and its spatial derivatives. This computation is
made in the Appendix \ref{Lcalculations}, where we obtain that the canonical
Hamiltonian density $\mathcal{H}$ (i.e., $H=\int d\mathbf{x}~\mathcal{H}$)
takes part in the results. These results are better understood when
projected on $E_{0}^{a}$ and $E_{k}^{a}$. Thus, we get the secondary
constraints written in canonical form:
\begin{equation}
G_{0}^{(2)}~\doteq ~\mathcal{H}-\partial _{i}(E_{0}^{c}~\Pi
_{c}^{i})~\approx ~0~,  \label{G20}
\end{equation}
\begin{equation}
G_{k}^{(2)}~\doteq ~\partial _{k}E_{i}^{c}~\Pi _{c}^{i}-\partial
_{i}(E_{k}^{c}~\Pi _{c}^{i})~\approx ~0  \label{G2K}
\end{equation}
(the symbol $\approx $ stands for equalities that are valid on the
constraint surface). The constraints (\ref{G20}) and (\ref{G2K}) are
equivalent to the super-Hamiltonian and super-momenta constraints of the
ADM\ formalism. While the ADM Hamiltonian vanishes on the constraint
surface, the TEGR Hamiltonian does not. The reason can be traced to the
surface term in Eq.~(\ref{RLC}); in fact, according to Eq.~(\ref{G20}), $
\mathcal{H}$ is not zero but a divergence (which became a spatial divergence
thanks to the constraints (\ref{G1a})).

\section{Gauge transformations}

\label{gauge_t}

We have already anticipated --although not proven yet-- that all the
constraints will be first class. So, let us consider the gauge
transformations generated by $G_{a}^{(1)}$ and $G_{\mu }^{(2)}$. In general,
the infinitesimal gauge transformation generated by a first class constraint
$G$ is \cite{Sundermeyer1982}
\begin{equation}
\delta E_{\mu }^{a}(t,\mathbf{x})\ =\ \int d\mathbf{y}~\epsilon (t,\mathbf{y}
)\ \{E_{\mu }^{a}(t,\mathbf{x}),G(t,\mathbf{y})\}~.  \label{delta}
\end{equation}
Any transformation of the vielbein has to be accompanied by a transformation
of the basis $\{\mathbf{e}_{a}\}$, in order to respect the duality relations
$\mathbf{E}^{a}(\mathbf{e}_{b})=\delta _{b}^{a}$ of Eq.~(\ref{duality}).
Therefore
\begin{equation}
\mathbf{E}^{a}(\delta \mathbf{e}_{b})+\delta \mathbf{E}^{a}(\mathbf{e}
_{b})~=~0~,  \label{deltae0}
\end{equation}
or
\begin{equation}
\delta e_{b}^{\nu }~=~-e_{a}^{\nu }~e_{b}^{\mu }~\delta E_{\mu }^{a}~.
\label{deltae}
\end{equation}
According to Eq.~(\ref{delta}), any linear combination of primary
constraints $\epsilon ^{b}(t,\mathbf{x})~G_{b}^{(1)}$ generates a
transformation that only affects the temporal component of the 1-forms $
\mathbf{E}^{a}$,
\begin{equation}
\delta E_{0}^{a}(t,\mathbf{x})~=~\epsilon ^{a}(t,\mathbf{x})~,~  \label{d1E}
\end{equation}
(or $\delta \mathbf{E}^{a}=\epsilon ^{a}dt$) which also implies \footnote{
Since $E=\varepsilon _{ab...g}\ E_{0}^{a}\ E_{1}^{b}\,...\,E_{n-1}^{g}$,
where $\varepsilon _{ab...g}$ is the Levi-Civita symbol, we also obtain $
e_{h}^{0}\,\delta E=e_{h}^{0}\ \epsilon ^{a}\ \varepsilon _{ab...g}\
E_{1}^{b}\,...\,E_{n-1}^{g}$ $=-E_{\nu }^{a}\ \delta e_{h}^{\nu }\
\varepsilon _{ab...g}\ E_{1}^{b}\,...\,E_{n-1}^{g}=-E\ \delta e_{h}^{0}$.
Therefore $e_{h}^{0}\,E$ is invariant under the transformation (\ref{d1E}).}
\begin{equation}
\delta e_{b}^{\nu }~=~-\epsilon ^{a}~e_{a}^{\nu }~e_{b}^{0}~.
\end{equation}
Instead, the transformations generated by $G_{0}^{(2)}$, $G_{k}^{(2)}$ only
affect the spatial components of the forms $\mathbf{E}^{a}$ (the canonical
Hamiltonian density $\mathcal{H}$ does not contain $\Pi _{a}^{0}$). Then,
the infinitesimal gauge transformations generated by $G_{0}^{(2)}$ and any
arbitrary combination $\xi ^{k}G_{k}^{(2)}$ are respectively
\begin{eqnarray}
\delta E_{i}^{a}(t,\mathbf{x}) &=&\xi \ \dot{E}_{i}^{a}(t,\mathbf{x}
)+E_{0}^{a}~\partial _{i}\xi   \notag \\
&=&\partial _{i}(E_{0}^{a}~\xi )+\xi \ 2~\partial _{\lbrack 0}E_{i]}^{a}~,
\label{d20E}
\end{eqnarray}
\begin{eqnarray}
\delta E_{i}^{a}(t,\mathbf{x}) &=&\xi ^{k}~\partial
_{k}E_{i}^{a}+E_{k}^{a}~\partial _{i}\xi ^{k}  \notag \\
\  &=&\partial _{i}(E_{k}^{a}~\xi ^{k})+\xi ^{k}~2~\partial _{\lbrack
k}E_{i]}^{a}~.  \label{d2kE}
\end{eqnarray}
In these results there is a term resembling the gauge transformation of the
electromagnetic potential. However, they come together with a term related
to the Weitzenb\"{o}ck torsion $\mathbf{T}^{a}=d\mathbf{E}^{a}$. Both terms
are needed because, differing from the electromagnetic Lagrangian, TEGR
Lagrangian depends not only on the exterior derivative of the field $\mathbf{
E}^{a}$ but on the field itself. Even so, the whole result exhibits a clear
geometric content, which can be evidenced by means of the Lie derivative of
a $p-$form $\mathbf{\alpha }$ along a vector $\mathbf{\xi }$,
\begin{equation}
\mathcal{L}_{\mathbf{\xi }}\mathbf{\alpha }~=~d[\mathbf{\alpha }(\mathbf{\xi
)}]+d\mathbf{\alpha }(\mathbf{\xi })~.  \label{Lie}
\end{equation}
In fact, the r.h.s. of Eqs.~(\ref{d20E}) and (\ref{d2kE}) constitute the
spatial components of $\mathcal{L}_{\mathbf{\xi }}\mathbf{E}^{a}$, where $
\mathbf{\xi }$ is the arbitrary vector field formed by the infinitesimal
parameters $\xi (t,\mathbf{x})$, $\xi ^{k}(t,\mathbf{x})$. We notice that
Eqs.~(\ref{d20E}) and (\ref{d2kE}) can be extended to the temporal component
of the 1-forms $\mathbf{E}^{a}$, since any change of $E_{0}^{a}$ is a gauge
transformation. Therefore, we have obtained that TEGR is insensitive to $2n$
independent gauge transformations of the vielbein on the constraint surface,
which are given by Eq.~(\ref{d1E}) and
\begin{equation}
\delta \mathbf{E}^{a}~=~\mathcal{L}_{\mathbf{\xi \,}}\mathbf{E}^{a}~.
\label{d2E}
\end{equation}
The derivative character of transformation (\ref{d2E}) together with Eq.~(
\ref{deltae0}) imply that
\begin{equation}
\delta \mathbf{e}_{b}~=~\mathcal{L}_{\mathbf{\xi \,}}\mathbf{e}_{b}~=~[
\mathbf{\xi \,},\mathbf{~e}_{b}]~.  \label{d2e}
\end{equation}
In turn, this last transformation leads to a change of the anholonomy
coefficients:
\begin{equation}
\delta f_{ab}^{c}~=~\mathcal{L}_{\mathbf{\xi \,}}f_{ab}^{c}~=~\mathbf{\xi }
(f_{ab}^{c})~,  \label{d2f}
\end{equation}
as can be easily verified by using the Jacobi identity to compute $\delta
\lbrack \mathbf{e}_{a}\mathbf{\,},~\mathbf{e}_{b}]~=~[\delta \mathbf{e}_{a}
\mathbf{\,},~\mathbf{e}_{b}]+[\mathbf{e}_{a}\mathbf{\,},~\delta \mathbf{e}
_{b}]$.

\bigskip

We remark that the Lie derivative of any Lagrangian --understood as the $n-$
form $\mathbf{L}~=~L~dx^{0}\wedge ...\wedge dx^{n-1}$, where $L$ is the
Lagrangian density-- is always a boundary term. In fact, if $\mathbf{\alpha }
$ is a $n-$form in Eq.~(\ref{Lie}), then its Lie derivative $\mathcal{L}_{
\mathbf{\xi }}\mathbf{\alpha }$ is the exact form $d[\mathbf{\alpha }(
\mathbf{\xi )}]$. But in a theory of gravity, like TEGR, this kind of
(quasi-) invariance of the Lagrangian comes from a symmetry of its dynamical
variables generated by a proper combination of the trivial primary
constraints and the secondary ones. In fact, the change of the TEGR
Lagrangian $n-$form,
\begin{eqnarray}
\mathbf{L}~ &=&~\dfrac{1}{8}~E~f_{ce}^{a}~f_{df}^{b}~M_{ab}^{\
\,\,\,cedf}~dx^{0}\wedge ...\wedge dx^{n-1}  \notag \\
&=&~\dfrac{1}{8}~f_{ce}^{a}~f_{df}^{b}~M_{ab}^{\ \,\,\,cedf}~\mathbf{E}
^{0}\wedge ...\wedge \mathbf{E}^{n-1}~,
\end{eqnarray}
(we used that the vielbein is orthonormal to rewrite the volume) under the
gauge transformation (\ref{d2E}) is equal to its Lie derivative by virtue of
Eqs.~(\ref{d2E}) and (\ref{d2f}):
\begin{eqnarray}
\delta \mathbf{L}~ &=&~\dfrac{1}{4}~\delta f_{ce}^{a}~f_{df}^{b}~M_{ab}^{\
\,\,\,cedf}~\mathbf{E}^{0}\wedge ...\wedge \mathbf{E}^{n-1}~  \notag \\
&&+\dfrac{1}{8}~f_{ce}^{a}~f_{df}^{b}~M_{ab}^{\ \,\,\,cedf}~\delta \mathbf{E}
^{0}\wedge ...\wedge \mathbf{E}^{n-1}+...  \notag \\
&=&~\mathcal{L}_{\mathbf{\xi \,}}\mathbf{L~=~}d[\mathbf{L}(\mathbf{\xi )}]~.
\end{eqnarray}

\section{More primary constraints. The Lorentz gauge group}

\label{lorentz_g}

So far we have found the $2n$ constraints that reflect the constraint
structure of the ADM formulation of general relativity. However, TEGR
describes the $n(n+1)$ components of the metric tensor through a $n\times n$
matrix $E_{\mu }^{a}$. The relation between both sets of dynamical variables
is given by the Eq.~(\ref{metric2}), which is invariant under local Lorentz
transformations of the vielbein. Since we know that TEGR has dynamics only
for the metric, as is clear from the equivalence between TEGR and
Einstein-Hilbert Lagrangians expressed in Eq.~(\ref{RLC}), the local Lorentz
symmetry has to be a property not only of the relation (\ref{metric2}) but
the set of dynamical equations. Then, we should find that Lorentz
transformations in the tangent space constitute a gauge group in TEGR.
Therefore, we will search for more primary constraints in Eq.~(\ref{momfull}
).

Eq.~(\ref{momfull}) is a system of $n^{2}$ equations that are not linearly
independent. In the previous Section we have already shown that they contain
a set of $n$ constraints that trivially emerge for $\mu =0$. The existence
of constraints associated to the temporal coordinate index is a consequence
of the privileged character the temporal coordinate plays in the canonical
formalism. We expect that the rest of the primary constraints are
exclusively related to tangent space indices. Therefore, we will look for
constraints among the coordinate invariant combinations $\Pi _{a}^{\mu
}E_{\mu }^{e}$; according to Eq.~(\ref{momfull}) they are
\begin{equation}
\Pi _{a}^{\mu }\,E_{\mu }^{e}\ =\ E\ C_{ab}^{\ \,\,\,\,ef}\,e_{f}^{\lambda
}\ \partial _{0}E_{\lambda }^{b}+E\ \partial _{i}E_{\lambda
}^{b}\,e_{c}^{0}\,e_{d}^{i}\,e_{f}^{\lambda }\,M_{ab}^{\ \,\,\,cedf},
\label{invariantmom}
\end{equation}
where $C_{ab}^{\ \,\,\,\,ef}$ is defined as
\begin{equation}
C_{ab}^{\ \,\,\,\,ef}~\doteq ~e_{c}^{0}\ e_{d}^{0}\ M_{ab}^{\ \,\,\,cedf}~.
\label{C}
\end{equation}
To find constraints (relations among the canonical variables) in Eq.~(\ref{invariantmom}), 
we should find (vielbein-depending) coefficients $
v_{\,\,e}^{a}$ such that $v_{\,\,e}^{a}\,\Pi _{a}^{\mu }E_{\mu }^{e}$ does
not contain canonical velocities. In other words, since the square matrix $
e_{f}^{\lambda }$ is not singular, it should be
\begin{equation}
v_{\,\,e}^{a}~C_{ab}^{\ \,\,\,\,ef}~=~0~.  \label{VC}
\end{equation}
Notice that even the $n$ trivial primary constraints $G_{g}^{(1)}\doteq \Pi
_{g}^{0}$ can be recovered in this way. In fact $G_{g}^{(1)}$ requires
coefficients $v_{|g|\,e}^{\,\,a}\doteq e_{e}^{0}~\delta _{g}^{a}$ (the index
between vertical bars is a label for each independent set of coefficients),
since $e_{e}^{0}~\delta _{g}^{a}~\Pi _{a}^{\mu }E_{\mu }^{e}=\Pi _{g}^{0}$.
On the other hand, these coefficients satisfy Eq.~(\ref{VC}), because $
M_{gb}^{\ \,\,\,cedf}$ is antisymmetric in $c-e$:
\begin{equation}
v_{|g|\,e}^{\,\,a}\ C_{ab}^{\
\,\,\,\,ef}~=~e_{e}^{0}~e_{c}^{0}~e_{d}^{0}~M_{gb}^{\ \,\,\,cedf}~\equiv ~0~.
\end{equation}

We will introduce an independent set of coefficients $v_{\,\,e}^{a}$ leading
to the primary constraints associated with the Lorentz group. Let be the set
of coefficients $v_{\,\,e}^{a}$ labeled by $gh$
\begin{equation}
v_{|gh|\,e}^{\ \ \ \,a}~\doteq ~2~\delta _{\lbrack g}^{a}~\eta _{h]e}~.
\end{equation}
Taking into account the form (\ref{supermetric}) of the supermetric, we
obtain
\begin{equation}
v_{|gh|\,e}^{\ \ \ \,a}\ C_{ab}^{\ \,\,\,\,ef}=2~e_{c}^{0}\ e_{d}^{0}\ \eta
_{e[h}\,M_{g]b}^{\ \,\,\,cedf}=4~e_{c}^{0}~e_{d}^{0}~\delta
_{hgb}^{cdf}\equiv 0,~  \label{VC2}
\end{equation}
since $\delta _{hgb}^{cdf}$ is completely antisymmetric (see Eq. \eqref{etaM} 
for details of the calculation). The antisymmetric labels $gh$
classify $n(n-1)/2$ new constraints. By combining Eqs.~(\ref{invariantmom}),
(\ref{VC2}) and (\eqref{etaM}), one gets
\begin{eqnarray}
0~ &\equiv &~v_{|gh|\,e}^{\ \ \ \,a}~(\Pi _{a}^{\mu }~E_{\mu
}^{e}-E~\partial _{i}E_{\lambda }^{b}\ e_{c}^{0}~e_{d}^{i}~e_{f}^{\lambda
}~M_{ab}^{\ \,\,\,cedf})~  \notag \\
&=&~2~\eta _{e[h}~\Pi _{g]}^{\mu }~E_{\mu }^{e}~+4~E~\partial _{i}E_{\lambda
}^{b}\ e_{[h}^{0}~e_{g}^{i}~e_{b]}^{\lambda }~.
\end{eqnarray}
In the last line, $\lambda $ can be replaced with $j$ due to the
antisymmetrization of the pair $h-b$. Besides, on the constraint surface it
is $\Pi _{g}^{0}=0$. So, we define the primary constraints
\begin{equation}
G_{gh}^{(1)}~\doteq ~2~\eta _{e[h}~\Pi _{g]}^{i}~E_{i}^{e}~+4~E~\partial
_{i}E_{j}^{b}\ e_{[h}^{0}~e_{g}^{i}~e_{b]}^{j}~\approx ~0~.  \label{LC}
\end{equation}
In Section \ref{full_brackets} we will prove that these $n(n-1)/2$\
constraints accomplish the Lorentz algebra. Besides, they will be consistent
with the evolution. The entire set of constraints will prove to be first
class. According to Eq.~(\ref{delta}), the gauge transformation of the
vielbein generated by a combination $\epsilon ^{gh}G_{gh}^{(1)}$ is
\begin{equation}
\delta E_{j}^{a}(t,\mathbf{x})=\int d\mathbf{y}~\epsilon ^{gh}(t,\mathbf{y}
)\,\{E_{j}^{a}(t,\mathbf{x}),2~\eta _{e[h}\,\Pi _{g]}^{i}(t,\mathbf{y}
)\,E_{i}^{e}\}~,
\end{equation}
which can be extended to the component $E_{0}^{a}$ by virtue of the gauge
transformation (\ref{d1E}), so leading to the local Lorentz transformation
\begin{equation}
\delta \mathbf{E}^{a}~=~\epsilon ^{gh}(t,\mathbf{x})~\left( \eta
_{eh}~\delta _{g}^{a}-\eta _{eg}~\delta _{h}^{a}\right) ~\mathbf{E}^{e}~.
\label{d3E}
\end{equation}

\bigskip

At this point, one could ask whether we have exhausted the solutions to Eq.~(
\ref{VC}). We remark that $C_{ab}^{\ \,\,\,\,ef}$ can be rephrased as a
symmetric $n^{2}\times n^{2}$ matrix by using a notation that take pairs of
flat indices $a,b,...$ to define a multi-index $A=\left( {}\right)
_{\,\,e}^{a}$ such that the Eq.~(\ref{VC}) becomes
\begin{equation}
v^{A}~C_{AB}~=~0  \label{eigen}
\end{equation}
For this, we use the following indexation formulas for $A=\left( {}\right)
_{\,\,e}^{a}$ , $B=\left( {}\right) _{\,\,f}^{b}$ \footnote{
The formula can be inverted by taking $a=[A/n]$, so $e=A-n[A/n]-1$, where $
[\ ]$ means the integer part.}
\begin{equation}
A=(a-1)\,n+e,\ \ \ \ B=(b-1)\,n+f;  \label{indexation}
\end{equation}
so, $A,B,...=1,...,n^{2}$. Eq.~(\ref{symM}) implies the symmetry of $C_{AB}$:
\begin{equation}
C_{AB}~=~C_{BA}~.  \label{Csym}
\end{equation}
Eq.~(\ref{eigen}) means that there are as many linear constraints as null
eigenvalues the symmetric $n^{2}\times n^{2}$ matrix $C_{AB}$ has. The
coefficients $v^{A}=v_{\,\,e}^{a}$\ of the constrained combinations $
v_{\,\,e}^{a}\,\Pi _{a}^{\mu }E_{\mu }^{e}$ are the components of the
respective eigenvectors. So far we have found $n+n(n-1)/2=n(n+1)/2$ null
eigenvalues. As we will see in the forthcoming sections, the other $n(n-1)/2$
eigenvalues are different from zero.

\section{TEGR canonical Hamiltonian}

\label{canonical_H}

We will fully exploit the multi-index notation introduced at the end of the
previous Section. For this, we define a set of objects of $n^{2}$
components:
\begin{eqnarray}
\dot{E}^{B} &\doteq &e_{f}^{\lambda }~\dot{E}_{\lambda }^{b}~,\hspace{0.2in}
E_{0}^{B}\doteq e_{f}^{i}~\partial _{i}E_{0}^{b}~,  \notag \\
\Pi _{A} &\doteq &\Pi _{a}^{\mu }\,E_{\mu }^{e}\ ,\hspace{0.2in}P_{A}\doteq
E\ \partial _{i}E_{k}^{b}~e_{c}^{0}~e_{d}^{i}~e_{f}^{k}~M_{ab}^{\
\,\,\,\,cedf}.~~~~
\end{eqnarray}

Thus the Lagrangian density~(\ref{LagrM}) reads
\begin{equation}
L=\frac{1}{2}~(\Pi _{A}+P_{A})(\dot{E}^{A}-E_{0}^{A})-U~,
\end{equation}
where
\begin{equation}
U\doteq -\frac{1}{2}\ E\ \partial _{i}E_{j}^{a}\ \partial _{k}E_{l}^{b}\
e_{c}^{i}\ e_{e}^{j}\ e_{d}^{k}\ e_{f}^{l}\ M_{ab}^{\;\;\;cedf}.
\end{equation}
Therefore, the canonical Hamiltonian density turns out to be
\begin{eqnarray}
\mathcal{H}\ &\doteq &\ \Pi _{a}^{\mu }~\dot{E}_{\mu }^{a}-L~=~\Pi _{A}\,\dot{E}
^{A}-L  \notag \\
&=&\frac{1}{2}\ (\Pi _{A}-P_{A})~\dot{E}^{A}+\frac{1}{2}~(\Pi
_{A}+P_{A})~E_{0}^{A}+U.~~~  \label{H}
\end{eqnarray}
To write $\mathcal{H}$ in a canonical way, the velocities $\dot{E}^{B}$ must
be solved in terms of the momenta. Eq.~(\ref{invariantmom}) displays the
linear relation among velocities and momenta; this equation now reads
\begin{equation}
\Pi _{A}-P_{A}\ =~E\ C_{AB}~(\dot{E}^{B}-E_{0}^{B})~.  \label{multipi}
\end{equation}
In Eq.~(\ref{multipi}) $\dot{E}^{B}$ cannot be straightforwardly solved
because the matrix $C_{AB}$ is singular. Matrix $C_{AB}$ has $n(n+1)/2$ null
eigenvalues, since there are $n(n+1)/2$ primary constraints linear in the
momenta. In spite of the fact that $C_{AB}$ is not invertible, we can still
solve the subspace of velocities that is orthogonal to the subspace of null
eigenvalues. In fact, by using a proper basis for splitting the subspace of
null eigenvalues, $C_{AB}$ would look like
\begin{equation}
C_{AB}^{\prime }=\left(
\begin{array}{cc}
0 & 0 \\
0 & \tilde{C}
\end{array}
\right) ,\hspace{0.1in}\
\end{equation}
In such a basis we would find $n(n+1)/2$ constraints $\Pi _{A}-P_{A}=0$;
besides, we would trivially solve $n(n-1)/2$ relevant velocities,
\begin{equation}
\dot{E}^{A}-E_{0}^{A}\ =~E^{-1}\ D^{\prime AB}~(\Pi _{B}-P_{B})~,
\label{solveE}
\end{equation}
where the matrix $D^{\prime }$ is
\begin{equation}
D^{\prime }=\left(
\begin{array}{cc}
0 & 0 \\
0 & \tilde{D}
\end{array}
\right) ,
\end{equation}
and satisfies
\begin{equation}
D^{\prime }~C^{\prime }=C^{\prime }~D^{\prime }=~\left(
\begin{array}{cc}
0 & 0 \\
0 & \mathbf{1}
\end{array}
\right) .  \label{1sub}
\end{equation}
Eq.~(\ref{solveE}) declares null the $n(n+1)/2$ first velocities. This
causes no harm, since these velocities enter the Hamiltonian (\ref{H}) as
the coefficients of the primary constraints $\Pi _{A}-P_{A}=0$. So, the
values of the $n(n+1)/2$ first velocities are irrelevant, because different
choices modify the Hamiltonian by terms proportional to the constraints.
Anyway this kind of terms are reintroduced in the primary Hamiltonian (\ref
{HP}).

Let us use the matrix $N$ of change of basis to return to the original
basis: $C^{\prime }=NCN^{-1}$. Then, the previous equation becomes
\begin{equation}
N^{-1}D^{\prime }N~C=C~N^{-1}D^{\prime }N=N^{-1}\left(
\begin{array}{cc}
0 & 0 \\
0 & \mathbf{1}
\end{array}
\right) N.
\end{equation}
The r.h.s. is not the identity, but is a symmetric matrix. Besides, the
matrix
\begin{equation}
D~\doteq ~N^{-1}D^{\prime }N  \label{D}
\end{equation}
satisfies that $CDC=C$ and $DCD=D$. Therefore $D$ is the
Moore-Penrose pseudo-inverse of $C$. We will use the
Eq.~(\ref{solveE}) in the original basis; so we must replace
$D^{\prime }$ with $D$. Thus, we substitute Eq.~(\ref{solveE}) in
Eq.~(\ref{H}) to obtain the canonical form of the
Hamiltonian density:
\begin{equation}
\mathcal{H}=\frac{1}{2}\ e\ (\Pi _{A}-P_{A})D^{AB}(\Pi _{B}-P_{B})+\Pi
_{A}\,E_{0}^{A}+U,  \label{canonicaldensityH}
\end{equation}
where $e = E^{-1} = \text{det}(e^{\mu}_a)$. The canonical Hamiltonian is the
integral of $\mathcal{H}$. We can remind the form (\ref{G20}) of the
constraint $G_{0}^{(2)}$ to write
\begin{equation}
H=\int ~d\mathbf{x}~\mathcal{H}=\int ~d\mathbf{x}~G_{0}^{(2)}+\int
E_{0}^{c}~\Pi _{c}^{i}~dS_{i}~.
\end{equation}
Then, the canonical Hamiltonian is a constraint plus a boundary term. As a
consequence, the set of first class constraints will be automatically
consistent with the evolution.

\subsection{\protect\bigskip Dimension $n=3$}

Let us work with the matrix $C_{\ B}^{A}$,
\begin{equation}
C_{\ B}^{A}=C_{\ eb}^{a\,\,\,\,f}=e_{c}^{0}\ e_{d}^{0}\ M_{\ b\ \,e}^{a\ g\
hf},  \label{C^A_B}
\end{equation}
where $M_{\ b\ \,e}^{a\ g\ hf}=\eta ^{ac}\ \eta _{de}\ M_{cb}^{\ \ gdhf}$ . $
C_{AB}$ and $C_{\ B}^{A}$ share the eigenvectors of null eigenvalue (see the
Appendix \ref{matrixC} for the forms of these matrices). The non-null
eigenvalues of $C_{\ B}^{A}$ are
\begin{eqnarray}
\lambda _{1} &=&\lambda
_{2}=2~[(e_{0}^{0})^{2}-(e_{1}^{0})^{2}-(e_{2}^{0})^{2}]=2~g^{00}\doteq
\lambda ~,  \notag \\
~\lambda _{3} &=&-\lambda ~.  \label{lambda3D}
\end{eqnarray}
The case $n=3$ is very simple because the matrix $C_{\ B}^{A}$ accomplishes
\begin{equation}
C_{\ B}^{A}~C_{\ C}^{B}~C_{\ D}^{C}~=\lambda ^{2}~C_{\ D}^{A}~.  \label{cubo}
\end{equation}
This is a consequence of the fact that the non-null eigenvalues have the
same absolute value. This means that the pseudo-inverse of $C_{\ B}^{A}$ is $
D_{\ B}^{A}=\lambda ^{-2}~C_{\ B}^{A}$. Therefore, the matrix $D^{AB}$ in
Eq.~(\ref{canonicaldensityH}) is
\begin{equation}
D^{AB}~=~\lambda ^{-2}~C^{AB}~=~\lambda ^{-2}~e_{g}^{0}\ e_{h}^{0}\ M_{\ \
\,\,\,\,e\ \,f}^{abg\ h}~.  \label{D3D}
\end{equation}

\subsection{Dimension $n>3$}

In $n=4$ dimensions, the matrix $C_{\ B}^{A}$ has six non-null eigenvalues;
they are
\begin{eqnarray}
\lambda _{1} &=&\lambda _{2}=\lambda _{3}=\lambda _{4}=\lambda _{5}  \notag
\\
&=&2~[(e_{0}^{0})^{2}-(e_{1}^{0})^{2}-(e_{2}^{0})^{2}-(e_{3}^{0})^{2}]=2~g^{00}\doteq \lambda ~,
\notag \\
\lambda _{6} &=&-2~\lambda ~.  \label{lambda4D}
\end{eqnarray}
Since their absolute values are not equal, the pseudo-inverse matrix $D^{AB}$
cannot be inferred in a so straightforward way as we did in $n=3$
dimensions. In fact, matrix $C_{\ B}^{A}$ does not accomplishes the Eq.~(\ref{cubo}) 
when $n>3$. The eigenvector related to the odd eigenvalue is
\begin{equation}
w^{B}=w_{\,f}^{b}=-\frac{\lambda }{2}~\delta _{f}^{b}+e_{f}^{0}\ \eta ^{bh}\
e_{h}^{0}~.  \label{W}
\end{equation}
In fact, in any dimension $n$, vector $w^{B}$ satisfies the eigenvalue
equation
\begin{eqnarray}
C_{\,\,B}^{A}~w^{B} &=&e_{g}^{0}\ e_{h}^{0}\ M_{\ b\ \,e}^{a\ g\
hf}w_{\,f}^{b}  \notag \\
&=&-(n-2)\lambda ~w_{\,e}^{a}=-(n-2)~\lambda ~w^{A}.  \label{Weigen}
\end{eqnarray}
We will show that the pseudo-inverse of $C_{\,\,B}^{A}$ can be formulated as
the matrix
\begin{equation}
D_{\,\,B}^{A}~=~\lambda ^{-2}~(C_{\,\,B}^{A}+\alpha ~w^{A}w_{B})~,
\label{DD}
\end{equation}
where $\alpha $ is a factor to be determined. The idea is to use the
projector associated to the odd eigenvalue to \textquotedblleft
improve\textquotedblright\ the matrix $C_{\,\,B}^{A}$ and get the desired
result. In order that $D_{\,\,B}^{A}$ be the pseudo-inverse of $
C_{\,\,B}^{A} $, the r.h.s. of the equation
\begin{equation}
C_{\,\,C}^{A}\,D_{\,\,D}^{C}\,C_{\,\,B}^{D}=\lambda
^{-2}C_{\,\,C}^{A}\,C_{\,\,D}^{C}\,C_{\,\,B}^{D}+\alpha (n-2)^{2}\,w^{A}w_{B}
\label{Calfa}
\end{equation}
should be $C_{\,\,B}^{A}$. To find $\alpha$, we will introduce the auxiliary
matrix
\begin{equation}
\tilde{C}_{\,\,B}^{A}=C_{\,\,B}^{A}+4~\lambda ^{-1}~w^{A}w_{B}~,
\label{Ctilde}
\end{equation}
which satisfies
\begin{eqnarray}
\tilde{C}_{\,\,B}^{A}~w^{B} &=&C_{\,\,B}^{A}~w^{B}+4~\lambda
^{-1}w^{A}\,w_{B}\,w^{B}~  \notag \\
&=&-(n-2)\,\lambda \,w^{A}+(n-1)\,\lambda \,w^{A}=\lambda \,w^{A}.\ \
\end{eqnarray}
Besides, for any vector $\ell ^{B}$ orthogonal to $w^{B}$ it is $\tilde{C}
_{\,\,B}^{A}~\ell ^{B}=C_{\,\,B}^{A}~\ell ^{B}=\lambda ~\ell ^{A}$ . Then, $
\tilde{C}_{\,\,B}^{A}$ is isotropic in the subspace of non-null eigenvalues.
\footnote{This is true not only for $n=4$, but it has been checked for arbitrary $n$
through the computer algebra program Cadabra \cite{Peeters:2007wn}.} Since
all the non-null eigenvalues of $\tilde{C}_{\,\,B}^{A}$ are equal to $
\lambda $, then $\tilde{C}_{\,\,B}^{A}$ accomplish the Eq.~(\ref{cubo}).
Therefore
\begin{eqnarray}
&&\lambda ^{2}~(C_{\,\,B}^{A}+4\,\lambda ^{-1}w^{A}w_{B})=\lambda ^{2}\,
\tilde{C}_{\,\,B}^{A}=\tilde{C}_{\,\,C}^{A}\,\tilde{C}_{\,\,D}^{C}\,\tilde{C}
_{\,\,B}^{D}  \notag \\
&=&C_{\,\,C}^{A}\,C_{\,\,D}^{C}\,C_{\,\,B}^{D}+4\,\lambda \left(
n^{2}-5n+7\right) w^{A}w_{B}~,
\end{eqnarray}
i.e.,
\begin{equation}
\lambda ^{-2}C_{\,\,C}^{A}C_{\,\,D}^{C}C_{\,\,B}^{D}=C_{\,\,B}^{A}-4~\lambda
^{-1}(n-3)(n-2)w^{A}w_{B}~.
\end{equation}
Substituting this result in Eq.~(\ref{Calfa}), we get that $D_{\,\,B}^{A}$
is the pseudo-inverse of $C_{\,\,B}^{A}$ if $\alpha $ has the value
\begin{equation}
\alpha \ =\ \lambda ^{-1}\ \frac{4~(n-3)}{(n-2)~}\ .  \label{alfa}
\end{equation}
In $n=4$ dimensions, $\alpha $ is equal to $2\lambda ^{-1}$. Thus the
contravariant pseudo-inverse matrix $D^{AB}=\lambda ^{-2}(C^{AB}+\alpha
~w^{A}w^{B})$\ in four dimensions is
\begin{eqnarray}
D^{AB}~ &=&D_{\ \ ef}^{ab}=\lambda ^{-1}~(\delta _{f}^{[a}~\delta _{e}^{b]}+
\frac{1}{2}\,\eta ^{ab}\,\eta _{ef})  \notag \\
&&-\lambda ^{-2}~(e_{e}^{0}\ e_{f}^{0}~\eta ^{ab}+4~e_{g}^{0}\
e_{[e}^{0}~\delta _{f]}^{[a}~\eta ^{b]g}  \notag \\
&&+e_{g}^{0}\ e_{h}^{0}~\eta ^{ag}\,\eta ^{bh}\,\eta _{ef})  \notag \\
&&+2~\lambda ^{-3}~\eta ^{ag}\,\eta
^{bh}~e_{g}^{0}~e_{h}^{0}~e_{e}^{0}~e_{f}^{0}~.  \label{DD4d}
\end{eqnarray}

\bigskip

\section{\protect\bigskip Algebra of constraints}

\label{full_brackets}

The Hamiltonian formalism for TEGR is not finished without checking that the
set of constraints is first class. For this, we have to compute the entire
set of Poisson brackets between the constraints. The pseudo-inverse matrix $
D^{AB}$ will enter the algebra of those Poisson brackets involving the
constraint $G_{0}^{(2)}$. It is worth mentioning that Eq.~(\ref{W}) can be
replaced in the l.h.s of Eq.~(\ref{Weigen}) to obtain
\begin{equation}
\frac{1}{2(n-2)}~C_{\ be}^{a\ \, b}\ =w_{\,\,e}^{a}=w^{A}.
\end{equation}
Therefore, matrix $D$ in Eq.~(\ref{DD}) can be written entirely in terms of
the matrix $C$ as
\begin{equation}
D_{\ \ ef}^{ab}=\lambda ^{-2}\,C_{\ \ ef}^{ab}+\dfrac{\lambda ^{-3}(n-3)}{
(n-2)^{3}}\,C_{\ ce}^{a\ \ c}\,C_{\ df}^{b\ \ d}\ ,
\end{equation}
which will be useful to compute those brackets involving $G_{0}^{(2)}$.

\bigskip The simplest brackets are those related to $G_{a}^{(1)}$:
\begin{eqnarray}
&&\{G_{a}^{(1)}(t,\mathbf{x}),G_{b}^{(1)}(t,\mathbf{y})\}=0,  \label{c1} \\
\medskip &&\{G_{i}^{(2)}(t,\mathbf{x}),G_{a}^{(1)}(t,\mathbf{y})\}=0,
\label{c2} \\
\medskip &&\{G_{ab}^{(1)}(t,\mathbf{x}),G_{c}^{(1)}(t,\mathbf{y})\}=0,
\label{c3} \\
&&\{G_{0}^{(2)}(t,\mathbf{x}),G_{a}^{(1)}(t,\mathbf{y})\}=  \notag \\
&&\ \ \ \left( e_{a}^{0}\,G_{0}^{(2)}+e_{a}^{i}\,G_{i}^{(2)}\right) \ \delta (
\mathbf{x}-\mathbf{y}).  \label{c4}
\end{eqnarray}
However, the last one requires the knowledge of the brackets between the
momenta $\Pi _{a}^{0}$ and the matrix $D^{AB}$. In the Appendix \ref{pbrackets} we
summarize useful hints in order to simplify this calculation.

The Poisson brackets between secondary constraints $G_{\mu }^{(2)}$
reproduce the algebra of constraints of the ADM formulation of general
relativity:
\begin{eqnarray}
\{G_{i}^{(2)}(t,\mathbf{x}),\,G_{j}^{(2)}(t,\mathbf{y})\} &=&\\ -G_{i}^{(2)}(
\mathbf{x})\ \partial _{j}^{\mathbf{y}}\delta
(\mathbf{x}-\mathbf{y})&+&G_{j}^{(2)}(\mathbf{y})\ \partial _{i}^{\mathbf{x}}\delta (\mathbf{x}-
\mathbf{y})\, ,\ \ \ \  \notag \label{c5} \\
\{G_{0}^{(2)}(t,\mathbf{x}),\,G_{0}^{(2)}(t,\mathbf{y})\} &=& g^{ij}(\mathbf{x})\, G^{(2)}_i(\mathbf{x})\, \partial^{\mathbf{y}}_j \delta(\mathbf{x}-\mathbf{y}) \notag \\
 &-& g^{ij}(\mathbf{y})\,G^{(2)}_i(\mathbf{y})\,\partial^{\mathbf{x}}_j\delta(\mathbf{x}-\mathbf{y})\,,  \ \ \ \ \ \ \label{c6} \\
\{G_{0}^{(2)}(t,\mathbf{x}),\,G_{i}^{(2)}(t,\mathbf{y})\} &=&G_{0}^{(2)}(
\mathbf{x})\ \partial _{i}^{\mathbf{y}}\delta
(\mathbf{x}-\mathbf{y})\, . \label{c7}
\end{eqnarray}
We have also verified that the Poisson brackets for the constraints
$G_{ab}^{(1)}$ reproduces the Lorentz algebra:
\begin{eqnarray}
\{&&G_{ac}^{(1)}(t,\mathbf{x}),\,G_{fe}^{(1)}(t,\mathbf{y})\}\ =  \label{c8} \\
&&\left( \eta _{ec}G_{af}^{(1)}+\eta _{af}G_{ce}^{(1)}-\eta
_{cf}G_{ae}^{(1)}-\eta _{ae}G_{cf}^{(1)}\right) \ \delta (\mathbf{x}-\mathbf{
y})\,.  \notag
\end{eqnarray}
Besides it is
\begin{equation}
\{G_{ab}^{(1)}(t,\mathbf{x}),\,G_{i}^{(2)}(t,\mathbf{y})\}=0\,.  \label{c9}
\end{equation}
Finally the most intricate calculation is required by the bracket
\begin{equation}
\{G_{0}^{(2)}(t,\mathbf{x}),\,G_{ab}^{(1)}(t,\mathbf{y})\}=E_{0}^{c}\,\eta
_{c[a}\,e_{b]}^{0}\,G_{0}^{(2)}\ \delta (\mathbf{x}-\mathbf{y})\,.
\label{c10}
\end{equation}
In order to alleviate some difficult parts of it, some useful computations
are summarized in Appendix \ref{pbrackets}.

As a result we have got $n$ trivial primary constraints $G^{(1)}_a$,
together with $n(n-1)/2$ primary constraints that come from the Lorentz
algebra. Besides, we have obtained $n$ secondary constraints $G^{(2)}_{\mu}$
that are equivalent to the super-Hamiltonian and super-momenta constraints
of the ADM formalism. Since we just proved that all constraints are first
class, then the counting of degrees of freedom goes as
\begin{eqnarray}
& \#\,\text{d.o.f.}\ =\ \#\,(\text{p,q}) -\#\,\text{f.c.c.}  \notag \\
& = n^2 - \dfrac{n(n+3)}{2} = \dfrac{n(n-3)}{2}
\end{eqnarray}
which is the number of degrees of freedom of general relativity in $n$
dimensions.

\section{Summary}

\label{conclusions} \bigskip

The essence of a Hamiltonian constrained system lies in the impossibility of
solving all the canonical velocities in terms of canonical momenta. This is
because the momenta are not independent but satisfy constraint equations,
which in turn means that some dynamical variables are spurious degrees of
freedom. In the case of the teleparallel equivalent of general relativity
(TEGR), such obstruction is expressed in the Eq.~(\ref{invariantmom}), since
$C_{ab}^{\ \,\,\,\,ef}$ cannot be inverted. $C_{ab}^{\ \,\,\,\,ef}$ is an
object intimately linked to the Lorentz invariant supermetric $M_{ab}^{\ \
cedf}$ entering the TEGR Lagrangian (\ref{Lagrf}). In order to analyze how
many constraints are involved in the Eq.~(\ref{invariantmom}), and how many
canonical velocities can be solved, we have arranged the components of $
C_{ab}^{\ \,\,\,\,ef}$ in a $n^{2}\times n^{2}$ symmetric matrix $C_{AB}$
(the relation between the superindex $A$ and the tangent space indices is given
in Eq.~(\ref{indexation})). We have shown that the eigenvalues of $
C_{\,\,B}^{A}$ follow a very simple pattern: $n(n+1)/2$ eigenvalues are
null, $n(n-1)/2-1$ of them are equal to $2\,g^{00}\doteq \lambda $, and the
remaining one is equal to $(2-n)\,\lambda $. The primary constraints results
from the contraction of the Eq.~(\ref{multipi}) with each eigenvector of
null eigenvalue; they include the $n$ trivial constraints $G_{a}^{(1)}$ (see
Eq.~(\ref{G1a})) and the $n(n-1)/2$ Lorentz constraints $G_{ab}^{(1)}$ (see
Eq.~(\ref{LC})). To build the canonical Hamiltonian we must identify the
subset of canonical velocities that can be still solved in terms of the
momenta. For this, we employed the Moore-Penrose pseudo-inverse of matrix $C$
, which can be sought in the form proposed in Eq.~(\ref{DD}) thanks to the
simple pattern of eigenvalues exhibited by the matrix $C$. The so obtained
matrix $D^{AB}$ is the piece we need to write the canonical Hamiltonian
density $\mathcal{H}$ (see Eq.~(\ref{canonicaldensityH})). Those terms
associated with the unsolved velocities are absorbed into the terms added to
the primary Hamiltonian $H_{p}$ (\ref{HP}). Besides the primary constraints $
G_{a}^{(1)}$, $G_{ab}^{(1)}$, we have also obtained $n$\ secondary
constraints $G_{\mu }^{(2)}$ --the diffeomorphism constraints-- that
guarantee that the primary constraints remain valid along the
evolution dictated by $H_{p}$ (we have examined this consistency at
the level of the Euler-Lagrange equations). The consistency under
the evolution of the system must be checked with the secondary
constraints too. Not surprisingly, the canonical Hamiltonian density
$\mathcal{H}$ is equal to $G_{0}^{(2)}$ except for a boundary term.
Thus, the consistency of the entire set of constraints is guaranteed
by the first class constraint algebra (\ref{c1}-\ref{c10}). Since
the constraints are first class, they generate gauge
transformations. Therefore, there are $n(n+3)/2$ spurious variables,
what reduces the number of degrees of freedom to $n(n-3)/2$. The
independent gauge transformations are those displayed in
Eqs.~(\ref{d1E}), (\ref{d2E}), and (\ref{d3E}).

\section*{Acknowledgments}

The authors thank N. Deruelle and C. Bejarano for helpful discussions. This
work was supported by Consejo Nacional de Investigaciones Cient\'ificas y
T\'ecnicas (CONICET) and Universidad de Buenos Aires.

\appendix

\section{}
\label{app}

\subsection{Properties of the supermetric}

\label{smprop}

There are many properties of the supermetric that were used throughout this
work, and can be deducted from its definition. Some of them are
\begin{equation}
M_{ab}^{\ \ cedf}=M_{ba}^{\ \ dfce}=-M_{ab}^{\ \ ecdf}=-M_{ab}^{\ \
cefd}.  \label{symM}
\end{equation}
We can calculate \textquotedblleft traces\textquotedblright\ of the
supermetric, which depend on the dimension $n$. Some of them are
\begin{eqnarray}
M_{ab}^{\ \ aedf}&=& M_{b a}^{\ \ d f a e} = 4(n-2)\,\eta ^{e[d}\,\delta_{b}^{f]}\,,\\
 M_{a b}^{\ \ d f a e} & = & M_{b a}^{\ \ a e d f} = 2(n-2) \eta^{e[f} \delta^{d]}_b, \\
M_{ab}^{\ \ aebf}&=&-2(n-1)(n-2)\,\eta ^{ef}\,,
\end{eqnarray}
The totally antisymmetric Kronecker delta $\delta _{cab}^{ghf}$
appears in the antisymmetrized product
\begin{eqnarray}
\eta _{e[c}M_{a]b}^{\ \ \, gehf} &=&2\, (\delta _{\lbrack
a}^{h}\delta _{c]}^{f}\delta _{b}^{g}+\delta _{\lbrack a}^{g}\delta
_{c]}^{h}\delta
_{b}^{f}+\delta _{\lbrack a}^{f}\delta _{c]}^{g}\delta _{b}^{h})  \notag \\
&\doteq &-2\ \delta _{cab}^{ghf}\ .  \label{etaM}
\end{eqnarray}
We also obtain
\begin{eqnarray}
M_{\ \ \ e\ f}^{abg\ h}\,\eta _{a[q}\delta _{p]}^{e}&=& \eta
^{bg}\eta _{f[q}\delta _{q]}^{h}+\eta ^{bh}\eta _{f[q}\delta
_{p]}^{g}+\delta
_{f}^{b}\delta _{\lbrack q}^{g}\delta _{p]}^{h}\, ,\ \ \ \ \ \\
\eta^{e[c}C^{d]b}_{\ \ ef} &=& 4\ e^0_f\ e^{0[c}\,\eta^{d]b} .
\end{eqnarray}
Some other combinations quadratic in $M$ appear in the calculations, and it
is useful to have them on hand
\begin{eqnarray}
C_{\ \ ec}^{ac}\, M_{a g}^{\ \ c e d f} &=& 4(n-3)(n-2)\,g^{00}\, \eta^{c[d}\, \delta^{f]}_{g}\ \ \ \ \ \ \ \ \ \ \ \ \ \ \\
&+& 8\,(n-2)\, e^{0}_g\, e^{0 [f}\, \eta^{d]c}\,+\, e^{0 c}\, e^{0 [d}\, \delta^{f]}_g \notag \\
C_{ab}^{\ \ ef}\, M_{\ \ \ e\ f}^{abc\ d} &=& 6\,(n-3)(n-2)\,\eta ^{cd}\,g^{00} \\
&+&12\,(n-2)\,e^{0c}\,e^{0d}\ .\notag
\end{eqnarray}

\subsection{Calculation of $\partial L/\partial E_{0}^{a}$}

\label{Lcalculations} For computing $\partial L/\partial E_{0}^{a}$, it is
important to notice that, contrarily to electromagnetism, $E_{0}^{a}$
appears in the Lagrangian not just in the spatial derivatives $\partial
_{i}E_{0}^{a}$ but also as a part of $e_{a}^{\mu }$ and $E$. First of all we
need the quotient $\partial e_{c}^{\mu }/\partial E_{\lambda }^{a}$, which
is obtained from the duality relation:
\begin{equation}
\delta _{\nu }^{\mu }=e_{b}^{\mu }E_{\nu }^{b}\ \ \ \rightarrow \ \ \ 0=
\dfrac{\partial e_{b}^{\mu }}{\partial E_{\lambda }^{a}}\,E_{\nu
}^{b}+e_{a}^{\mu }\,\delta _{\nu }^{\lambda }\ .
\end{equation}
this implies that
\begin{equation}
\dfrac{\partial e_{c}^{\mu }}{\partial E_{\lambda }^{a}}=-e_{a}^{\mu
}\,e_{c}^{\lambda }\ .
\end{equation}
We will need also the expression $\partial E/\partial E_{0}^{a}$,
which is obtained from the explicit formula for the determinant
\begin{equation}
E=\epsilon
_{abcd...g}\,E_{0}^{a}\,E_{1}^{b}\,E_{2}^{c}\,E_{3}^{d}\,...E_{n}^{g}\,,
\end{equation}
Then we obtain
\begin{equation}
E_{\lambda }^{a}\,\dfrac{\partial E}{\partial E_{0}^{a}}=\delta _{\lambda
}^{0}\,E\ \ \ \rightarrow \ \ \ \dfrac{\partial E}{\partial E_{0}^{a}}
=E\,e_{a}^{0}\ .
\end{equation}
In this way,
\begin{eqnarray}
&&\dfrac{\partial L}{\partial E_{0}^{a}} =\dfrac{1}{2}\,E\,(e_{a}^{0}e_{g}^{
\mu }e_{e}^{\nu }e_{h}^{\rho }e_{f}^{\lambda }-e_{a}^{\mu
}e_{g}^{0}e_{e}^{\nu }e_{h}^{\rho }e_{f}^{\lambda }-e_{a}^{\nu }e_{g}^{\mu
}e_{e}^{0}e_{h}^{\rho }e_{f}^{\lambda }  \notag \\
&&-e_{a}^{\rho }e_{g}^{\mu }e_{e}^{\nu }e_{h}^{0}e_{f}^{\lambda
}-e_{a}^{\lambda }e_{g}^{\mu }e_{e}^{\nu }e_{h}^{\rho
}e_{f}^{0})\,\partial _{\mu }E_{\nu }^{c}\,\partial _{\rho
}E_{\lambda }^{d}\,M_{cd}^{\ \ gehf}.\ \ \ \
\end{eqnarray}
In the last expression we identify the Lagrangian in the first term, and
different index combinations of the momenta. We rewrite it and continue with
the algebraic manipulation
\begin{eqnarray}
\dfrac{\partial L}{\partial E_{0}^{a}} &=&e_{a}^{0}\,L-\dfrac{1}{2}
e_{a}^{\mu }\,\partial _{\mu }E_{\nu }^{c}\,\Pi _{c}^{\nu }+\dfrac{1}{2}
\,e_{a}^{\nu }\,\partial _{\mu }E_{\nu }^{c}\,\Pi _{c}^{\mu }  \notag \\
&&-\dfrac{1}{2}\,e_{a}^{\rho }\,\partial _{\rho }E_{\lambda }^{d}\,\Pi
_{d}^{\lambda }+\dfrac{1}{2}\,e_{a}^{\lambda }\,\partial _{\rho }E_{\lambda
}^{d}\,\Pi _{d}^{\rho }  \notag \\
&=&e_{a}^{0}\,L+2\,e_{a}^{\nu }\,\partial _{\lbrack \mu }E_{\nu ]}^{c}\,\Pi
_{c}^{\mu } \\
&=&e_{a}^{0}\,L+2\,e_{a}^{0}\,\partial _{\lbrack i}E_{0]}^{c}\,\Pi
_{c}^{i}+2\,e_{a}^{j}\,\partial _{\lbrack i}E_{j]}^{c}\,\Pi _{c}^{i}\,.
\notag
\end{eqnarray}
The Hamiltonian density can be extracted from the first terms, to
obtain
\begin{eqnarray}
\dfrac{\partial L}{\partial E_{0}^{a}} &=&-e_{a}^{0}\,\mathcal{H}
+e_{a}^{0}\,\partial _{i}E_{0}^{c}\,\Pi _{c}^{i}+2\,e_{a}^{j}\,\partial
_{\lbrack i}E_{j]}^{c}\,\Pi _{c}^{i} \\
&=&e_{a}^{0}\,(\partial _{i}(E_{0}^{c}\Pi _{c}^{i})-\mathcal{H}
)-E_{0}^{c}\,e_{a}^{0}\,\partial _{i}\Pi _{c}^{i}+2\,e_{a}^{j}\,\partial
_{\lbrack i}E_{j]}^{c}\,\Pi _{c}^{i}\,.  \notag
\end{eqnarray}
This result is replaced in Eq.~(\ref{secondary}) to obtain $n$
secondary constraints:
\begin{equation}
E_{j}^{c}\,e_{a}^{j}\,\partial _{i}\Pi _{c}^{i}+e_{a}^{0}\,(\partial
_{i}(E_{0}^{c}\Pi _{c}^{i})-\mathcal{H})+2\,e_{a}^{j}\,\partial _{\lbrack
i}E_{j]}^{c}\,\Pi _{c}^{i}\approx 0\,.
\end{equation}
We note that only spatial derivatives are present, and the canonical
Hamiltonian takes part in the secondary constraints. We can isolate the
contribution of the Hamiltonian by doing the contraction with $E_{0}^{a}$;
thus we get
\begin{equation}
G_{0}^{(2)}=\mathcal{H}-\partial _{i}(E_{0}^{c}\,\Pi _{c}^{i})\approx 0\,.
\end{equation}
Besides, we perform the contraction with $E_{k}^{a}$, so yielding
\begin{equation}
G_{k}^{(2)}=\partial _{k}E_{i}^{c}\,\Pi _{c}^{i}-\partial
_{i}(E_{k}^{c}\,\Pi _{c}^{i})\approx 0\,.
\end{equation}

\subsection{Matrix $C^A_{\ B}$}

\label{matrixC}

We present the full expression for the matrix $C^{A}_{\ B}$ in $n=4$, which
appears in the definition of the canonical momenta. It is
\begin{widetext}
\begin{footnotesize}
\begin{equation}
C^A_{\ B}=
\left(
\begin{tabular}{cccccccccccccccc}
\smallskip
 0 & 0 & 0 & 0 & 0 & $2c_{23}$ & $ -2d_{12}$ & $-2d_{13}$ & 0 & $-2d_{12}$ & $2c_{13}$ & $-2d_{23}$ & 0 & $-2d_{13}$ & $-2d_{23}$ & $2c_{12}$ \\
 \smallskip
 0 & $-c_{23}$ & $d_{12}$ & $d_{13}$ & $c_{23}$ & 0 & $-d_{02}$ & $ -d_{03}$ & $-d_{12}$ & $-d_{02}$ & $2d_{01}$ & 0 & $-d_{13}$ & $-d_{03}$ & 0 & $2d_{01}$ \\
 \smallskip
 0 & $ d_{12}$ & $-c_{13}$ & $d_{23}$ & $-d_{12}$ & $2d_{02}$ & $-d_{01}$ & $0$ & $c_{13}$ & $-d_{01}$ & $0$ & $-d_{03}$ & $-d_{23}$ & $0$ & $-d_{03}$ & $2d_{02}$ \\
 \smallskip
 0 & $ d_{13}$ & $d_{23}$ &  $-c_{12}$ & $-d_{13}$ & $2d_{03}$ & $0$ &  $-d_{01}$ & $-d_{23}$ & $0$ & $2 d_{03}$ &  $-d_{02}$ & $c_{12}$ & $-d_{01}$ & $-d_{02}$ &  $0$ \\
 \smallskip
 0 & $c_{23}$ & $-d_{12}$ & $-d_{13}$ & $-c_{23}$ & $0$ & $d_{02}$ & $d_{03}$ & $d_{12}$ & $d_{02}$ & $-2d_{01}$ & $0$ & $d_{13}$ & $d_{03}$ & $0$ & $-2d_{01}$ \\
 \smallskip
 $2c_{23}$ & 0 & $-2d_{02}$ & $-2d_{03}$ & $0$ & $0$ & $0$ & $0$ & $2d_{02}$ & $0$ & $-2c_{03}$ & $-2d_{23}$ & $2d_{03}$ & $0$ & $-2d_{23}$ & $-2c_{02}$ \\
 \smallskip
 $-2 d_{12}$ & $d_{02}$ & $d_{01}$ & $0$ & $-d_{02}$ & $0$ & $c_{03}$ & $d_{23}$ & $-d_{01}$ & $c_{03}$ & $0$ & $d_{13}$ & $0$ & $d_{23}$ & $d_{13}$ & $-2d_{12}$ \\
 \smallskip
 $-2 d_{13}$ & $d_{03}$ & $0$ & $d_{01}$ & $-d_{03}$ & $0$ & $d_{23}$ & $c_{02}$ & $0$ & $d_{23}$ & $-2d_{13}$ & $d_{12}$ & $-d_{01}$ & $c_{02}$ & $d_{12}$ & $0$\\
 \smallskip
 0 & $-d_{12}$ & $c_{13}$ & $-d_{23}$ & $d_{12}$ & $-2d_{02}$ & $d_{01}$ & $0$ & $-c_{13}$ & $d_{01}$ & $0$ & $d_{03}$ & $d_{23}$ & $0$ & $d_{03}$ & $-2d_{02}$ \\
 \smallskip
 $-2 d_{12}$ & $d_{02}$ & $d_{01}$ & $0$ & $-d_{02}$ & $0$ & $c_{03}$ & $d_{23}$ & $-d_{01}$ & $c_{02}$ & $0$ & $d_{13}$ & $0$ & $d_{23}$ & $d_{13}$ & $-2d_{12}$ \\
 \smallskip
 $2c_{13}$ & $-2d_{01}$ & $0$ & $-2d_{03}$ & $2d_{01}$ & $-2c_{03}$ & $0$ & $-2d_{13}$ & $0$ & $0$ & $0$ & $0$ & $2d_{03}$ & $-2d_{13}$ & $0$ & $-2c_{01}$ \\
 \smallskip
 $-2 d_{23}$ & 0 & $d_{03}$ & $d_{02}$ & $0$ & $-2d_{23}$ & $d_{13}$ & $d_{12}$ & $-d_{03}$ & $d_{13}$ & $0$ & $c_{01}$ & $-d_{02}$ & $d_{12}$ & $c_{01}$ & $0$ \\
 \smallskip
 0 & $-d_{13}$ & $-d_{23}$ & $c_{12}$ & $d_{13}$ & $-2d_{03}$ & $0$ & $d_{01}$ & $d_{23}$ & $0$ & $-2d_{03}$ & $d_{02}$ & $-c_{12}$ & $d_{01}$ & $d_{02}$ & $0$ \\
 \smallskip
 $-2 d_{13}$ & $d_{03}$ & $0$ & $d_{01}$ & $-d_{03}$ & $0$ & $d_{23}$ & $c_{02}$ & $0$ & $d_{23}$ & $-2d_{13}$ & $d_{12}$ & $-d_{01}$ & $c_{02}$ & $d_{12}$ & $0$ \\
 \smallskip
 $-2 d_{23}$ & 0 & $d_{03}$ & $d_{02}$ & $0$ & $-2d_{23}$ & $d_{13}$ & $d_{12}$ & $-d_{03}$ & $d_{13}$ & $0$ & $c_{01}$ & $-d_{02}$ & $d_{12}$ & $c_{01}$ & $0$ \\
 \smallskip
 $2c_{12}$ & $-2 d_{01}$ & $-2d_{02}$ & $0$ & $2d_{01}$ & $-2c_{02}$ & $-2d_{12}$ & $0$ & $2d_{02}$ & $-2d_{12}$ & $-2c_{01}$ & $0$ & $0$ & $0$ & $0$ & $0$ \\
 \end{tabular}
\right)
\end{equation}
\end{footnotesize}
\end{widetext}
where
\begin{equation}
\begin{split}
& c_{01}=(e^{0}_0)^2-(e^{0}_1)^2 , \ \ \ c_{02}=(e^{0}_0)^2-(e^{0}_2)^2, \\
& c_{03}=(e^{0}_0)^2-(e^{0}_3)^2 , \ \ \ c_{12}=(e^{0}_1)^2+(e^{0}_2)^2 , \\
& c_{13}=(e^{0}_1)^2+(e^{0}_3)^2 , \ \ \ c_{23}=(e^{0}_2)^2+(e^{0}_3)^2 , \\
& d_{01}=e^0_0 e^0_1 , \ \ \ d_{02}=e^0_0 e^0_2, \ \ \ d_{03}=e^0_0 e^0_3, \\
& d_{12}=e^0_1 e^0_2 , \ \ \ d_{13}=e^0_1 e^0_3 , \ \ \ d_{23}=e^0_2 e^0_3.
\end{split}
\end{equation}
The matrices $C_{AB}$ and $C^{AB}$ are obtained by raising and lowering
indices with the corresponding $\eta $ tensors. The matrix $D_{\ B}^{A}$ is
obtained starting from (\ref{DD}).

\subsection{Poisson brackets}

\label{pbrackets}

Some useful fundamental Poisson brackets between the canonical variables and
their derivatives are given below,
\begin{eqnarray}
\{E(t,\mathbf{x}),\,\Pi _{a}^{\mu }(t,\mathbf{y})\}\ &=&\ E\ e_{a}^{\mu }\ \delta (
\mathbf{x}-\mathbf{y}),\\
\{e(t,\mathbf{x}),\,\Pi _{a}^{\mu }(t,\mathbf{y})\}\ &=&\ e\ e_{a}^{\mu }\ \delta (
\mathbf{x}-\mathbf{y}),
\end{eqnarray}
\begin{eqnarray}
\{e_{a}^{\mu }(t,\mathbf{x}),~\Pi _{b}^{\nu
}(t,\mathbf{y})\}~&=&~-e_{b}^{\mu }~e_{a}^{\nu }~\delta
(\mathbf{x}-\mathbf{y})\,,\ \ \ \ \ \ \
\\ \notag\\ \{\partial _{\lambda }E_{\mu
}^{a}(t,\mathbf{x}),~\Pi _{b}^{\nu }(t,\mathbf{y})\} &=&\\ -\{E_{\mu
}^{a}(t,\mathbf{x}),~\partial _{\lambda }\Pi _{b}^{\nu }(t,
\mathbf{y})\} \ &=&\ \delta _{b}^{a}~\delta _{\mu }^{\nu }~\partial
_{\lambda }^{\mathbf{x} }\delta (\mathbf{x}-\mathbf{y})\, ,\notag
\end{eqnarray}
\begin{equation}
\{\partial _{\mu }E_{\nu }^{b}(t,\mathbf{x}),\partial _{\lambda }\Pi
_{c}^{\lambda }(t,\mathbf{y})\}=\int d\mathbf{z}\ \delta _{c}^{b}\,\partial
_{\mu }^{x}\delta (\mathbf{x}-\mathbf{z})\,\partial _{\nu }^{y}\delta (
\mathbf{y}-\mathbf{z}).
\end{equation}
These expressions are enough (together with patience and a lot of calculations) 
to calculate those Poisson brackets that do not involve $G_{0}^{(2)}$. 
For the remaining Poisson brackets, we provide some easy-to-derive expressions
\begin{eqnarray}
\{G_{0}^{(2)}&&(t,\mathbf{x}),\, E_{i}^{c}(t,\mathbf{y)}\} = \\
&&\left(e\, D^{AB}(\Pi _{B}-P_{B})\, E_{i}^{e}\ \delta_{a}^{c} + \partial _{i}E_{0}^{c}\right)\, 
\delta(\mathbf{x}-\mathbf{y})\, ,\ \ \notag \\
\{G_{0}^{(2)}&&(t,\mathbf{x}),\, \partial _{\lambda }E_{\mu}^{c}(t,\mathbf{y})\} = \\
&&\left(e\,D^{AB}\,(\Pi _{B}-P_{B})\ E_{\mu }^{e}\ \delta _{a}^{c}
+\partial _{\mu }^{x}E_{0}^{c}(x)\right)\, \partial _{\lambda}^{y}\delta (\mathbf{x} -\mathbf{y})\, .  \notag
\end{eqnarray}
Other combinations of brackets between canonical momenta and some
basic building blocks of the secondary constraint $G_{0}^{(2)}$,
that recurrently appear in the calculations, are the following
\begin{eqnarray}
\{\lambda ^{-\gamma},\,\Pi _{c}^{0}\} &=& \ 2\ \gamma\
\lambda^{-\gamma}\ e_{c}^{0}\,,
\label{LPi} \\
\{\lambda^{-\gamma},\,\Pi _{c}^{i}\} &=&\ 4\ \gamma\, \lambda
^{-(\gamma+1)}\ e_{c}^{0}\ g^{0i}\,,
\\
\{w^{A},\,\Pi_{c}^{0}\} &=&-2\ e_{c}^{0}\ w^{A}\,,
\\
\{w^{A},\,\Pi _{c}^{i}\}
&=&-\dfrac{1}{2(n-2)}\,e_{c}^{0}\,(e_{g}^{i}
\,e_{h}^{0}+e_{h}^{i}\,e_{g}^{0})\,M_{\ d\ e}^{a\ g\ hd}.\ \ \ \ \ \ \
\end{eqnarray}
Finally, we give some help to calculate the brackets of the momenta
and the matrix $D^{AB}$. It is very simple to get the brackets
\begin{equation}
\{D^{AB},\Pi _{c}^{0}\}=2\ e_{c}^{0}\ D^{AB}.
\end{equation}
However for the spatial part of the momenta $\Pi _{a}^{i}$, the
brackets with the matrix $D$ do not simplify so easily. After using
all the developed tools, we get
\begin{eqnarray}
&&\{D^{AB},\,\Pi
_{c}^{i}\}=8\,e_{c}^{0}\,g^{0i}\ \lambda ^{-3}\,
\left( C^{AB} + \alpha \,w^{A}w^{B}\right) \notag \\
&&-\lambda^{-2}\, e^0_c \, (e_{g}^{i}e_{h}^{0}+e_{h}^{i}e_{g}^{0})\,
M_{\ \ \ \, e\ f}^{abg\ h}  \notag \\
&&-\dfrac{\alpha\, \lambda^{-2}\ e^0_c }{2(n-2)} \,
(e_{g}^{i} e_{h}^{0} + e_{h}^{i} e_{g}^{0})\, ( M^{a\ g\ hd}_{\ d\ e}w^B +M^{b\ g\ hd}_{\ d\ f} w^A ) \notag \\
&&+ 4\,\alpha\, \lambda^{-3}\, e^0_c \,  g^{0i}\,  w^A\, w^B .
\label{DPi}
\end{eqnarray}

In Eqs.~(\ref{LPi}-\ref{DPi}) a factor $\delta (\mathbf{x}
-\mathbf{y})$ is understood. As a general advice, the raising and
lowering of indices in the supermetric $M_{ab}^{\ \ cedf}$ must be
carefully done, to keep the original order of the indices and
protect the symmetries of the object. 

\end{document}